\documentstyle[preprint,prl,aps]{revtex} 
\newcommand{\be}{\begin{equation}} 
\newcommand{\ee}{\end{equation}}

\begin{document} 
\title{Apparent Rate Constant for Diffusion-Controlled Threemolecular reaction} 
\author{S.F. Burlatsky\dag} 
\author{M. Moreau} 
\address{LPTL, Universit\'e Paris VI, 4 Place Jussieu, 75252 Paris, France\\ 
\dag {\em Pemanent address:\/} Institute of \\ 
Chemical Physics, \\ 
Russian Academy of Sciences, Moscow 117977, Russia} 
\date{} 
\maketitle 
 
\begin{abstract} 
We present simple explicit estimates for the apparent reaction rate constant 
for threemolecular reactions. For small concentrations and $d> 1$, it 
depends only on the diffusion coefficients and sizes of the particles. For 
small concentrations and $d\le 1$, it is also time -- dependent. For large 
concentrations, it gains the dependence on concentrations. 
\end{abstract} 
 
 
 
\section{Introduction} 
 
\indent 
Threemolecular processes are important for catalysis \cite{catal}, where two 
particles, say $A$ and $B$, react in presence of the third substance, 
``catalytic cite'' --- $C$. Possible applications to coagulation were also 
concerned \cite{Krapicoag}. 
 
To our knowledge, the first attempt to obtain a simple theoretical 
description of the kinetics of the threemolecular reaction\footnote{ 
The {\em Product} can contain $C$ and/or $A$ or $B$.}  
\begin{equation}  \label{TMR} 
A+B+C\stackrel{k_t}{\Longrightarrow }{\em Product}, 
\end{equation} 
where $k_t$ is the ``chemical'' reaction rate constant, which determines the 
reaction rate per one threeparticle encounter, was made in \cite 
{Burlthree,BurlKD}, where the heuristic mean-field like analysis of the 
reaction kinetics for large $C$ --- sites was presented. Recently this 
subject gained more interest \cite{OshBlThre,Krapicoag,PrivGrinb,BenabrThre}. 
 
The theory of the {\em bimolecular\/} diffusion-controlled reactions  
\begin{equation}  \label{BMR} 
A+B\stackrel{k_b}{\Longrightarrow } {\em Product} 
\end{equation} 
started with the mean-field type approach in $d=3$ \cite{Smoluch}. The most 
important result of this theory is the fundamental Smoluchowsky reaction 
rate constant, $k_{Smol}$, which determines the apparent rate constant 
$k_{app}$ and, therefore, the mean reaction rate  
\begin{equation}  \label{arr} 
\frac{dC_A}{dt}=-k_{app}C_AC_B, 
\end{equation} 
where $C_A$ and $C_B$ are the mean concentrations. Frequently $k_{app}
$obeys the ``inverse resistance law''\cite{MorStoch1,MorStoch2}  
\begin{equation}  \label{irl} 
\frac{1}{k_{app}}=\frac{1}{k_{transp}}+\frac{1}{k}_{chem}. 
\end{equation} 
 
For the reaction (\ref{BMR}) in $3d$ the ``transport'' constant 
$k_{transp}=k_{Smol}$ and the ``chemical'' constant $k_{chem}=k_b$. First 
attempts to improve the Smoluchowsky theory concerned corrections to 
$k_{Smol}$ and it took a long time, starting from the early works \cite 
{BalVak,Burlcor,ZeldOvch,TV,RK,MeakStan}, to recognize that fluctuation 
effects, which determine the long-time asymptotic for the concentration 
dependencies, can be important. Contrary to it, in threemolecular reactions 
theory, the most of the recent works are concentrated on the long time 
behavior (See, however, \cite{OshBlThre} where the whole time domain was 
studied by means of a new elegant method)
and/or $d=1$ in spite of the fact that complete mean-field like theory is 
not finished and the analog of $k_{Smol}$ was not calculated. Particularly,  
\cite{OshBlThre,Krapicoag,BenabrThre,PrivGrinb} stress that for the long-time kinetics 
of the particular type of the reaction (\ref{TMR}), 
with $A\equiv B\equiv C$ ,
 i.e.\ for $A+A+A\stackrel{k_t}{\Longrightarrow }lA$ with $l<3$, the 
fluctuation effects are not decisive and that the conce
ntration of $A$, $C_A$, for $d>1$ is governed by  
\begin{equation}  \label{MFL} 
C_A\propto 1/{\sqrt t}, 
\end{equation} 
and for $d=1$ contains ``logarithmic corrections''  
\begin{equation}  \label{MFLlog} 
C_A\propto \sqrt{(\ln t)/t}. 
\end{equation} 
 
In our paper we present simple estimates for the Smoluchowsky - like rate 
constant, $k_{th}$, which determines for threemolecular reactions the 
apparent rate constant in the equation  
\begin{equation}  \label{AML} 
\frac{d C_A(t)}{d t}=-k_{app}\delta _A C_A(t)C_B(t)C_C(t), 
\end{equation} 
where $\delta _A =$ the number of $A$ particles, which enter the reaction, 
$minus$ the number of $A$ particles which the {\em Product} contains,  
for all $d$. We show that: ({\em i}) For small $C$ --- sites Eq.~(\ref{AML}) 
leads to the dependence Eq.~(\ref{MFL}) for $d>1$. For $d=1$ we regain the 
result, Eq~(\ref{MFLlog}), since $k_{th}$ becomes time dependent.  
We stress that in this case the result is still mean field one, since the 
fluctuations are not incorporated (see \cite{BurlDD} for more detailed 
analysis of the fluctuation effects on the k - particle reactions with 
different types of initial distributions). However, even for $d>1$,
 $k_{app}$ is renormalized by the transport processes and, therefore, it is not equal 
to $k_t$. ({\em ii}) For large $C$ --- sites  
$k_{app}$ in the Eq.~(\ref{AML}) becomes a function of the mean 
concentrations and, threfore, Eqs.~(\ref{MFL},\ref{MFLlog}) are not valid. 
 
In order to understand the main scaling laws for $k_{th}$ we consider first 
the simplified hoping picture with correlations in the reaction zone and 
without correlations in larger distance. For important limiting cases we 
present also the more traditional, diffusion - reaction equation approach. 
To make clear the ideas we start with the bimolecular reaction (\ref{BMR}). 
 
\section{Bimolecular reactions} 
 
\subsection{Smoluchowsky theory} 
 
\label{brsmolth} According to the Smoluchowsky theory, the concentration of 
$B$ particles in the distance $r$ from the center of the $A$ particle, 
$C_B(r,t)$, is governed by the diffusion equation  
\begin{equation}  \label{Deq} 
\frac{C_{B}(r,t)}{dt}=D_{AB}\nabla ^2C_B(r,t), 
\end{equation} 
where $D_{AB}=D_A+D_B$, with the boundary conditions  
\begin{equation}  \label{bcr} 
\Phi\equiv S_dD_{AB}\nabla C_B(r,t)\left|_{r= R_{AB}} 
=k_bC_{B}(R_{AB},t),\right. 
\end{equation} 
\begin{equation}  \label{bci} 
C_{B}(r=\infty ,t)=C_B, 
\end{equation} 
where $R_{AB}=R_{A}+R_{B}$ and $\Phi$ is a flux of $B$ particles through the  
$d$ --- dimensional sphere, $|r|=R_{AB}$, with the surface $S_d$. The 
reaction rate is equal to the quasi steady state value of $\Phi$ multiplied 
by $C_A$. For $d=3$ it leads to the Eq.~(\ref{irl})
 with $k_{chem}=k_b$
 and $k_{transp}=k_{Smol}$,  
\begin{equation} 
k_{Smol}= \left\{  
\begin{array}{ll} 
4\pi RD_{AB} & \mbox{for $d=3$} \label{sc3} \\  
\frac{2\pi D_{AB}}{\ln \frac{R^2}{D_{AB}t}}\label{sc2} & \mbox{for $d=2$} \\  
\sqrt{{D_{AB}}/{\pi t}} \label{sc1} & \mbox{for $d=1$}. 
\end{array} 
\right. 
\end{equation} 
For $t\gg R_{AB}^2/D$ in $d<3$ and for $k_b\gg k_{Smol}$ for $d\ge 3$, 
$k_{app}= k_{Smol}$. In $d=3$, the deviation of $C_B(r,t)$ from the limiting 
value, $C_B$, decreases proportional to ${R_{AB}}/{|r|}$ and, therefore, the 
size, $L$, of the ``reaction zone'',
 where the distributions of $A$ and $B$ are correlated,
 is of the order of the reaction radius $R_{AB}$. Contrary to  
$d=3$, in $d=1$ the correlated region grows proportionally to $\sqrt{D_{AB}t} 
$. The $d=2$ case is marginal. For smaller dimensions, $d\leq 2$, the 
diffusion is recurrent, the space exploration is compact \cite{Gennesreact}, 
which means that the number of the returns of the diffusing particle to the 
origin tends to infinity if $t\rightarrow \infty $. Therofore, the volume of 
the reaction zone equals to \cite{Gennesreact}  
\begin{equation}  \label{rvol} 
\Omega \propto L^d, 
\end{equation} 
where  
\begin{equation}  \label{ldif3} 
L \propto R_{AB}\ \mbox{for $d > 2$}\ \mbox{and}\ L \propto \sqrt{D_{AB}t}\  
\mbox{for $d \leq 2$}. 
\end{equation} 
 
\subsection{Hoping model} 
 
\label{brhm} 
 
Let us now consider the simplified picture of the reaction (\ref{BMR}), 
where the correlations for $A$ and $B$ are present only in the reaction zone 
and the exchange of $A(B)$ particles between the reaction zone and the non 
disturbed region is a ``one step'' process with the frequency, $\nu 
_{A(B)}\propto D_{A(B)}/L^2$, which equals to the minimum eigenvalue for the 
corresponding diffusion problem. In this section, we also assume that the 
concentration of the reacting particles is small, so we can consider only 
pairs $AB$ and neglect the configurations with more particles $B(A)$ in the 
reaction zone of $A(B)$. Therefore, the reaction rate is proportional to the 
number of $AB$ pairs, $N_{AB}$, and the balance equation reads  
\begin{equation}  \label{1balance} 
\frac{dN_{AB}}{dt}=\nu _{AB}\left( \Omega VC_AC_B-N_{AB}\right) -k_b\frac{ 
N_{AB}}{\Omega},  \label{diffterm} 
\end{equation} 
where $V$ is the total volume of the system and $\nu _{AB}=\nu _A+\nu _B$. 
The first term in the r.h.s.\ of Eq.~(\theequation) is the rate of jumps 
into the reaction zone, the second term is is the rate of jumps from the 
reaction zone and the third term is the reaction rate. The quasi steady 
state solution of Eq.~(\ref{1balance}) for the reaction rate leads to the 
ressult pesented by Eq.~(\ref{arr}) and Eq.~(\ref{irl})
 with $k_{chem}=k_b$ 
and $k_{transp}=\nu _{AB}\Omega $.  
Taking advantage of the Eqs.~(\ref{sc3})--(\ref{ldif3}), we conclude that  
\begin{equation} 
\nu _{AB}\Omega =k_{Smol}  \label{bhrr} 
\end{equation} 
with the accuracy of the insignificant numerical multipliers for $d=1$ and 
$d\geq 3$, and with the accuracy of $\propto \log (t)$ corrections for $d=2$. 
 
\section{Threemolecular reactions.} 
 
\label{tr}  
 
\subsection{Probability distributions.} 
 
\label{prob} Consider the reaction (\ref{TMR}) with $D_C=0$. In order to 
take thremolecular correlations into account we write down the master 
equation for the joint probability, $P(n_A,n_B)$, to obtain $n_A$ particles 
$A$ and $n_B$ particles $B$ in the reaction zone, near a $C$---site,  
\begin{equation}  \label{probrate} 
\frac{dP(n_A,n_B)}{dt}=I_{diff}+I_{react}. 
\end{equation} 
The diffusion and reaction terms are equal to  
\begin{eqnarray}  \label{pr3} 
I_{diff} & =& -\left(\nu _An_A+\nu _Bn_B+\nu _A\Omega C_A+\nu _B\Omega 
C_B\right)P(n_A,n_B)+  \nonumber \\ 
& & (n_A+1)\nu _AP(n_A+1,n_B)+(n_B+1)\nu _BP(n_A,n_B+1)  \nonumber \\ 
& & +\mbox{}\nu_A\Omega C_AP(n_A-1,n_B)+\nu_B\Omega C_BP(n_A,n_B-1), 
\end{eqnarray} 
\begin{eqnarray}  \label{prate3} 
I_{react}& =&-R(n_A,n_B)P(n_A,n_B)+  \nonumber \\ 
& & R(n_A+1,n_B+1)P(n_A+1,n_B+1), 
\end{eqnarray} 
where $R(n_A,n_B)$ is the reaction rate for the reaction zone with the given 
numbers of $A$ and $B$.   
We assume that the local reaction rate is equal to
 the product of $A$ and $B$ 
concentrations in the reaction zone,  
\begin{equation} 
R(n_A,n_B)=k_t\frac{n_An_B}{\Omega ^{2}}.  \label{rrate} 
\end{equation} 
 
\subsection{Hoping model} 
 
\label{trhm} 
 
\subsubsection{Small A and B concentrations (small C --- sites)} 
 
\label{trhmsc} \label{introd} For small A and B concentrations, 
$C_A(C_B)\Omega<<1$, we consider only triples, the particles $C$, which have 
$A$ {\em and\/} $B$ in the reaction zone, and pairs, the particles $C$ which 
have $A$ {\em or\/} $B$ in the reaction zone. 
The mean number of $C$ 
particles is $N_C=C_CV$; of $CA$ pares --- $N_{CA}=N_CP\{1,0\}$; 
and of $CB$ 
pares --- $N_{CB}=N_CP\{0,1\}$. The reaction rate is proportional to the 
number of the triples, $N_{CAB}=N_CP\{1,1\}$. From the Eqs.~(\ref{probrate} 
)-(\ref{rrate}) we obtain    
\begin{eqnarray}  \label{tdiffterm} 
\frac{dN_{CA(B)}}{dt}&=&\nu _A(\Omega VC_CC_{A(B)}-N_{CA(B)})-  \nonumber \\ 
& & \nu _B(\Omega N_{CA(B)}C_{B(A)} +N_{CAB}), \\ 
\frac{d N_{CBA}}{d t}&=&\nu _B(\Omega VC_CC_B-N_{CB})+  \nonumber \\ 
& &\nu_A(\Omega VC_CC_A-N_{CA})- k_t\Omega ^{-2}N_{CBA}. \label{nnn}
\end{eqnarray}

When $\nu _A>>\nu _B\Omega C_B$ and $\nu _B>>\nu _A\Omega C_A$,  
the steady state state solution of the Eqs.~(\ref 
{tdiffterm}),(\ref{nnn})  leads to the Eq.~(\ref{AML}) for the 
reaction rate and to Eq.~(\ref{irl}) 
for $k_{app}$ with $k_{chem}=k_t$ 
and $k_{transp}= \nu 
_{AB}\Omega ^2$.  
Taking advantage of the Eqs.~(\ref{sc3})--(\ref{ldif3}),  
we conclude that with the accuracy of the logarithmic 
corrections for $d=1$ 
the threemolecular analog of the Smoluchowsky constant equals to\footnote{ 
Note, however, that in this case the reaction on the catalytic surface with 
the typical time $\tau _{surf}^{-1} \propto 
R_{A,B}C_{A,B}^{(2)}(D_{A}+D_{B}) $ can become the limiting stage of the 
reaction.} 
\begin{equation}  \label{thsmd} 
k_{th}\propto \nu _{AB}\Omega ^2 \propto R^{2d-2}D_{AB}. 
\end{equation}

When one of the diffusion coefficients, say $D_B$, 
tends to zero, Eqs.~(\ref{tdiffterm}),(\ref{nnn}) lead to  
\begin{equation}  \label{smallD} 
\frac{dC_A}{dt}=-\nu_B\Omega \delta _AC_C C_B\propto -k_{Smol}\delta _AC_C 
C_B. 
\end{equation} 
 
\subsubsection{Large A or B concentrations (large C --- sites)} 
 
\label{trhmlc} 
 
For $C_{A(B)}\Omega >>1$, the analysis of the Eqs. (\ref{probrate})-(\ref 
{rrate}) for the averaged $A(B)$ concentrations in the reaction zone, 
$C_{A(B)}^{(r)}\equiv \sum^\infty _{n_A=0}\!\sum^\infty _{n_B=0} 
P(n_A,n_B)n_{A(B)}/\Omega$, leads to  
\begin{equation}  \label{cCA(B)diffterm} 
\frac{dC_{A(B)}^{(r)}}{dt}=\nu_A(C_{A(B)}-C_{A(B)}^{(r)})-k_t\frac{ 
C_A^{(r)}C_B^{(r)}}{\Omega}. 
\end{equation} 
Note that the reaction term (the last term in the right hand side of the 
Eq.~(\ref{cCA(B)diffterm})) in this regime {\em decouples into the product 
of the concentrations}. For the quasi steady state Eq.~(\ref{cCA(B)diffterm} 
) reduces to an algebraic second order equation for the reaction rate.  
When $k_tC_B<<\nu _A\Omega$, it leads to the Eq.~(\ref{AML}) with  
\begin{equation}  \label{keffh} 
k_{app}=\frac{\nu _A\Omega k_t}{\nu _A\Omega +k_tC_A}. 
\end{equation} 
When $k_tC_B>>\nu _A\Omega$ and $k_tC_A>>\nu _B\Omega$, the steady state Eq. 
(\ref{cCA(B)diffterm}) leads to  
\begin{equation}  \label{2limlargeh} 
\frac{dC_A}{dt}=\Omega C_C\min (\nu _AC_A,\nu _BC_B). 
\end{equation}

\subsection{Diffusion appoach} 
 
\label{trda} \label{DA}  
 
\subsubsection{Small A and B concentrations (small C---sites)} 
 
\label{SCL} For the small concentration limit we propose the following 
extension of the Smoluchowsky approach to threemolecular reactions. We place 
the origin into the center of an immovable particle $C$ and determine the 
the conditional density of triples, 
$C_{AB}(r_A,r_B,t)d^dr_Ad^dr_B$=$ 
N_{CAB}(r_A,r_B,t)/(VC_C)$, where $N_{CAB}(r_A,r_B,t)$ is the number of 
triples with $A$ in the volume element $d^dr_A$ near 
the end of the $d$ 
dimensional radius - vector $dr_A$ {\bf and} $B$ in the volume element 
$d^dr_B$ near the end of the $d$ dimensional radius - vector $dr_B$. The 
analog of Eq.~(\ref{Deq}) reads  
\begin{equation} 
\frac{\partial C_{AB}(r_A,r_B,t)}{\partial t}=\left( D_A\nabla 
_{r_A}^2+D_B\nabla _{r_B}^2\right) C_{AB}\left( r_A,r_B,t\right) . 
\label{3molsmol} 
\end{equation} 
For the spherically symmetric system, i.e. for $D_A=D_B=D/2$ and $C_A=C_B$,  
Eq.~(\ref{3molsmol}) reduces to  
\begin{equation} 
\frac{\partial C_{AB}(r,t)}{\partial t}=Dr^{1-2d}\frac \partial {\partial r} 
\left( r^{2d-1}\frac{\partial C_{AB}(r,t)}{\partial r}\right)  
\label{3molsmolspher} 
\end{equation} 
with the following boundary conditions  
\begin{equation} 
\lim C_{AB}(r,t)_{r\rightarrow \infty }=C_AC_B,  \label{infbc} 
\end{equation} 
\begin{equation} 
D_{AB}S_{2d}\left. \frac{\partial C_{AB}(r,t)}{\partial r}\right| 
_{r=R}=k_tC_{AB}(R,t),  \label{3molsmolspherbc} 
\end{equation} 
where $r=\sqrt{r_A^2+r_B^2}$. The solution of the Eq. (\ref{3molsmolspher}) 
with the boundary conditions (\ref{infbc}), (\ref{3molsmolspherbc}) leads to 
the Eq. (\ref{thsmd}) for $d>1$. For $d=1$.  
\begin{equation} 
k_{th}\propto \frac{D_{AB}}{\ln \frac{R^2}{D_{AB}t}}.  \label{thsmd1} 
\end{equation} 
 
\subsubsection{Large $A$ and $B$ concentrations (large $C$ --- sites)} 
 
\label{LCO} Choosing the origin in the center of the $C$ particle, we write 
the following decoupled equations for $A$ and $B$ concentrations  
\begin{equation}  \label{BOA} 
\frac{\partial C_{A(B)}(r,t)}{\partial t}=D_{A(B)}\nabla ^2C_{A(B)}(r,t) 
\end{equation} 
with the boundary conditions  
\begin{equation}  \label{BOAbc} 
D_{A(B)}S_d\frac{\partial C_{A(B)}(r,t)}{\partial r}\bigg | 
_{=R_{CA(B)}}=k_tC_A(r,t)C_B(r,t). 
\end{equation} 
When $k_tC_B<<4\pi R_{CA}D_A$, the solution of the Eqs. (\ref{BOA}--\ref 
{BOAbc}) leads to the Eq.~(\ref{AML}) for the reaction rate with  
\begin{equation}  \label{keff} 
k_{app}=\frac{k_{Smol}k_t}{k_{Smol}+k_tC_A}. 
\end{equation} 
When $k_tC_B>>4\pi R_{CA}D_A$ and $k_tC_A>>4\pi R_{CB}D_B$, it leads to 
unusual result  
\begin{equation}  \label{2limlarge} 
\frac{dC_A}{dt}=4\pi C_C\min (C_AR_{CA}D_A,C_BR_{CB}D_B). 
\end{equation} 
 
\section{Discussion} 
 
Thus, the hoping and diffusion models lead to the similar results for 
apparent rate constants --- for bimolecular reaction: Eqs.~(\ref{sc3}) and ( 
\ref{bhrr}) and for threemolecular reactions: Eqs.~(\ref{thsmd}, \ref{smallD} 
) and Eqs.~(\ref{thsmd1}) --- for small $C$ -- sites; and Eqs.~(\ref{keffh},  
\ref{2limlargeh})) 
and Eq.~(\ref{keff}, \ref{2limlarge}) --- for large $C$ 
-- sites. The diffusion model provides more precise results for the marginal 
dimensions: $d=2$ for bimolecular reactions, Eq.~(\ref{sc3}), and $d=1$ for 
threemolecular reactions, Eq.~(\ref{thsmd1}). On the other hand, the hoping 
model provides an easier way for the analysis of the correlations structure, 
Section \ref{prob}, and for small diffusion coefficient limit, Eq.~(\ref 
{smallD}). 
 
In small $C$ --- limit the threemolecular $ABC$ correlations are important, 
since the reaction act --- annihilation of a triple --- changes the 
concentrations of both $A$ and $B$ in the reaction zone from it`s maximum 
value to zero, 
i.e.\ $\delta (C^r_A)\equiv \delta (C^r_B) \equiv C^r_{A(B)}$ 
. At the same time, since the concentrations are small, the linear boundary 
condition, Eq.~(\ref{3molsmolspherbc}), is valid. The reaction occurs, when  
$A$ particle joins the $CB$ pair or  
when $B$ particle joins the $CA$ pair. 
Therefore, the limiting reaction rate, Eqs.~(\ref{thsmd}, \ref{thsmd1}), in 
this case equals to the rate of jumps of the particles A(B) to the reaction 
zone of $CB(A)$ pair, multiplied by the concentration of CB(A) pairs, which 
meens that $k_{th} \propto \Omega k_{Smol}$. When $D_B \rightarrow 0$, the 
reaction is limited by the transport of $B$ particles to $C$ and the 
reaction rate, Eq.~(\ref{smallD}), is the same as for the bimolecular 
reaction $B+C\Rightarrow 0$ with $k_{app} \Rightarrow k_{Smol}$. 
 
For large $C$ --- sites fluctuations of $A$ and $B$ concentrations in the 
reaction zone are much smaller than the mean values and, therefore, the 
decoupled (but non linear) equations for $A$ and $B$ concentrations, Eqs.~( 
\ref{cCA(B)diffterm}) and (\ref{BOA}, \ref{BOAbc}), are valid. For $k_t 
\rightarrow 0$, Eqs.~(\ref{keffh}, \ref{keff}) predict non renormalized 
value, $k_{app} \approx k_t$; for $k_t \rightarrow \infty$, Eqs.~(\ref{keffh} 
, \ref{2limlargeh}, \ref{keff}, \ref{2limlarge}) predict the same reaction 
rate as for diffusion- controlled bimolecular reaction. Note that the 
reaction ``chooses'' from $A$ and $B$ the reagent with the minimum effective 
reaction rate.\footnote{ 
An interesting problem concerns the fluctuation induced kinetics in the 
system with $C_AR_{CA}D_A=C_BR_{CB}D_B$.} More detailed and rigorous 
analysis of the correlator structure will be published in our subsequent 
publication. 
 
\acknowledgements{The authors thank Profs. A. Blumen, G.S. Oshanin, A.A. 
Ovchinnikov and S. Redner for stimulating discussions.}  
\bibliographystyle{prsty} 
\bibliography{coto}

\end{document}